\documentclass[fleqn,10pt]{elsarticle}
\usepackage{amssymb}
\usepackage{amsmath}
\usepackage{relsize}
\usepackage{subfig}
\usepackage{graphicx}
\usepackage{color}
\usepackage{multirow}
\usepackage{wasysym}
\usepackage[english]{babel}
\usepackage[symbol]{footmisc}
\usepackage{url}

\journal{Plos One}

\begin{document}

\title{Impact and Recovery Process of Mini Flash Crashes: An Empirical Study}

\date{\today}

\author[ude]{T.~Braun\corref{cor1}\tnoteref{t1}}

\author[ude]{J.A.~Fiegen\tnoteref{t1}}

\author[ude]{D.C.~Wagner}

\author[ude]{S.~Krause}

\author[ude]{T.~Guhr}

\cortext[cor1]{Corresponding authors: tobias.braun.13@stud.uni-due.de \\
jonas.fiegen@uni-due.de}
\tnotetext[t1]{Equally contributed}

\address[ude]{Faculty of Physics, University of Duisburg-Essen, Lotharstrasse 1, 47057 Duisburg, Germany\vskip 18pt \textnormal{\today}}


\begin{abstract}
In an Ultrafast Extreme Event (or \textit{Mini Flash Crash}), the price of a traded stock increases or decreases strongly within milliseconds. We present a detailed study of Ultrafast Extreme Events in stock market data. In contrast to popular belief, our analysis suggests that most of the Ultrafast Extreme Events are not primarily due to High Frequency Trading. In at least 60 percent of the observed Ultrafast Extreme Events, the main cause for the events are large market orders.
In times of financial crisis, large market orders are more likely which can be linked to the significant increase of Ultrafast Extreme Events occurrences.
Furthermore, we analyze the 100 trades following each Ultrafast Extreme Events. While we observe a tendency of the prices to partially recover, less than 40 percent recover completely. On the other hand we find 25 percent of the Ultrafast Extreme Events to be almost recovered after only one trade which differs from the usually found price impact of market orders.
\end{abstract}

\flushbottom
\maketitle
\thispagestyle{empty}

\section{Introduction}

Within the last two decades, algorithmic trading gained in importance at global stock markets~\cite{Breuer2012,Handershott2013}. In contrast to conventional traders, algorithmic traders automatically make trading decisions, place and observe orders~\cite{Hendershott2009}. The algorithms that are used remain secret simply because front--running~\cite{Harris1997} needs to be avoided. Consequently, not much literature is available about trading algorithms. This increasing influence of algorithmic trading within the past years prompted the installation of the ATP--flag (Automated Trader Program) as an indicator of algorithmic trading at the Xetra stock exchange~\cite{Xetra2004}. Currently, an equivalent tool indicating algorithmic trading is not available at US markets. There are clear differences in human versus algorithmic trading behaviour. One obvious advantage of algorithmic traders over humans is the quicker reaction time, which for example can be exploited for high frequency arbitrage~\cite{Foucault}. The drastically increasing performance of networks and computers during the past decades accelerates this progress~\cite{Moore1965,supercomputer,Young1997}. Not surprisingly, High Frequency Trading (HFT) is often being criticized by market participants as well as by the media and in the political discussion~\cite{schneller,english}.

During the rise of algorithmic trading, a new challenge emerged in the form of so called flash crashes, with large price changes in very short times \cite{haldane2012race,Johnson2013}. The flash crash of May 6 in 2010 produced the largest ever intraday point decline of Dow Jones of almost 1000 points. This event affected many stocks, some of them loosing almost their entire value, others rising by more than a factor of one thousand \cite{haldane2012race}. As the shock appeared too fast for human intervention, some market transactions have been rescinded later that day, and later market regulation established upper bounds for price movements in short periods \cite{serritella2010high}.

There is also a large number of less drastic mini flash crashes, usually in only one stock at a time \cite{nanex, golub2012high, nokermantaxonomy}. The reasons of flash crashes are discussed controversially. Explanations include imbalanced liquidity \cite{easley2011microstructure, andersen2014vpin, cespa2014illiquidity}, market manipulation \cite{brush2015mystery}, and large orders are discussed as triggers, from mistyped fat finger trades to intermarket sweep orders \cite{golub2012high}. A special focus is on HFT \cite{kirilenko2011flash,johnson2012financial,Johnson2013,pan2013can}, especially because on short time scales algorithmic feedbacks could be dominating, without possibility for human intervention. 
Johnson \textit{et al.}\cite{Johnson2013} find $18\,520$ Ultrafast Extreme Events (UEEs) using a criterion described below. They regard HFT as a potential cause for financial crises~\cite{faz2013}.
On the other hand, it has been pointed out that such price jumps without incoming new information are typical for price processes in general \cite{filimonov2012quantifying, hardiman2013critical}, even including human participation.

Another open question is the impact of flash crashes. While the prices after the big flash crash of 2010 recovered rapidly \cite{madhavan2012exchange}, the crash still demonstrated extreme risks for investors which are hard to evaluate and to handle \cite{haldane2012race}. The fast price recovery of flash crashes implies that they can be considered as bursts of volatility rather than as price jumps \cite{christensen2014fact}. Anyhow, the price recovery after mini flash crashes has so far not been systematically evaluated. A comparison of recovery times with the usual price impact of market orders \cite{bouchaud2004fluctuations,Bouchaud200957,toth2011anomalous} could be interesting.

In an attempt to contribute to a clarification of these issues, we have a closer look on the interplay between UEEs and HFT by using order flow data~\cite{Friedman1993,Johnson2010}. We test how often single trades dominate the flash crashes. If this happens often, most flash crashes occur without an algorithmic feedback as discussed in ~\cite{Johnson2013}. Furthermore we analyze, how often the price is restored to levels close to the price before the flash crash, and compare our findings with the common price impact of market orders \cite{bouchaud2004fluctuations}. 
The paper is organized as follows. We present our statistical analysis in \textit{Results}. We discuss which mechanisms trigger UEES in \textit{Mechanism for UEEs}. Finally, we analyze the impact of UEEs on the further development of traded prices in \textit{Impact and recovery of UEEs}.
We conclude our findings in \textit{Discussion}.

\section{Results}
In \textit{Data set and general statistical properties} we introduce our data set and the criteria used to identify UEEs. Moreover we focus on statistical aspects of UEEs. When did they occur predominantly? Which stocks and stock exchanges are most effected? Do these events occur simultaneously across different stocks and stock exchanges? What is the typical size of these events?
 
\subsection{Data set and general statistical properties}
\label{sec2}
We consider a data set of trades and quotes for all stocks of the S\&P 500 which were continously traded during 2007 and 2008. 
The data set was acquired from the NYSE Group \cite{NYX} (\grqq TAQ Database Release 3.0\grqq ) and contains data with a timestamp precision of one second on the following stock exchanges: American Stock Exchange (today: NYSE MKT), Archipelago Exchange (today: NYSE Arca), BATS Global Markets, Boston Stock Exchange (today: NASDAQ OMX BX), Chicago Stock Exchange, International Securities Exchange, NASDAQ Stock Market, NASDAQ ADF (Alternative Display Facility), National Stock Exchange, New York Stock Exchange and Philadelphia Stock Exchange (today: NASDAQ OMX PHLX).
To deal with the limited timestamp precision, we arranged trades with the same timestamp equidistantly within the second they occur.

The trades data offer sufficient information for detecting the occurrence of UEEs. Those that are detected on larger time scales such as so called \emph{breakdowns}\cite{Gao} are excluded in this analysis. Here we use the commonly employed criterion put forward by Nanex~\cite{nanex}, where an UEE occurs whenever the traded price changes monotonously by at least 0.8 percent within $1.5$ seconds and at least ten trades. Thereby we refer to a flash crash (spike) when the price moves in negative (positive) direction.

Following the above mentioned UEE criterion we find 5529 UEEs in our dataset. Table~\ref{tab:ueesec} groups them according to the industrial sectors. It is remarkable that the financial sector clearly dominates with $33.35$ events per company on average, followed by the energy sector with $14.36$ and the telecommunication service sector with $12.14$. In addition, the standard deviation is extremely high for firms in the financial sector: for instance, the stock of Morgan Stanley (MS) exhibits 717 such UEEs during 2007 and 2008.
Furthermore Table~\ref{tab:uee1s} shows that there are seconds in time in which more than one UEE occurs. In all of these cases the UEEs occur in different stocks within the same seconds. For example, the probability that more than one UEE occurs within a second is about $9\,\%$. At 10:35:01 on the 10th of December 2008 eight UEEs occured within one second on different stocks and stock exchanges which were APC (Anadarko Petroleum Corp.) on New York Stock Exchange, BHI (Baker Hughes Inc.) on BATS Global Markets, CAM (Cameron International Corp.) on BATS Global Markets, NE (Noble Corp.) on Archipelago Exchange, NOV (National Oilwell Varco Inc.) on Archipelago Exchange as well as VLO (Valero Energy) on New York Stock Exchange, Archipelago Exchange und NASDAQ Stock Market. All involved companies belong to the energy sector and all eight UEEs were flash crashes.

\begin{table}[htbp]
\centering
\begin{tabular}{ccccc}
\multirow{2}{*}{\textbf{\begin{tabular}[c]{@{}c@{}}industrial sector\\(GICS)\end{tabular}}} & \multirow{2}{*}{\textbf{\begin{tabular}[c]{@{}c@{}}number of\\companies\end{tabular}}} & \multicolumn{3}{c}{\textbf{number of events}}                                                                            \\ \cline{3-5} 
                                                                                                             &                                                                                            & \textbf{total} & \textbf{\begin{tabular}[c]{@{}c@{}}per\\ company\end{tabular}} &\textbf{\begin{tabular}[c]{@{}c@{}}standard\\ deviation\end{tabular}} \\ \hline
Consumer Discretionary                   & 79                                                                                         & 508            & 6.43		& 5.76                                                                    \\ \hline
Consumer Staples                                & 38                                                                                         & 122            & 3.21   	& 2.69                                                                    \\ \hline
Energy                                                 & 39                                                                                         & 560            & 14.36  	& 23.08                                                                   \\ \hline
Financials                                            & 77                                                                                         & 2568           & 33.35    & 84.51                                                                   \\ \hline
Health Care                                         & 48                                                                                         & 249            & 5.19     & 5.75                                                                    \\ \hline
Industrials                                          & 61                                                                                         & 274            & 4.49     & 7.27                                                                    \\ \hline
Information Technology                 & 65                                                                                         & 486            & 7.48     & 8.35                                                                    \\ \hline
Materials                                            & 31                                                                                         & 362            & 11.68    & 19.49                                                                   \\ \hline
Telecommunications Services                  & 7                                                                                          & 85             & 12.14    & 12.51                                                                   \\ \hline
Utilities                                           & 33                                                                                         & 315            & 9.55     & 42.14                                                                   \\ \hline
\end{tabular}
\caption{Occurrences of UEEs depending on their industrial sector (as per \textbf{G}lobal \textbf{I}ndustry \textbf{C}lassification \textbf{S}tandard). For every sector we calculate the total and average number of occurrences as well as the corresponding standard deviation.}
\label{tab:ueesec}
\end{table}

\begin{table}[htbp]
\centering
\begin{tabular}{|c|c|c|c|c|c|c|c|c|}
\hline
\textbf{UEEs within one second} & 1 & 2 & 3 & 4 & 5 & 6 & 7 & 8\\ \hline
\textbf{occurrences} & 4514 & 372 & 50 & 16 & 6 & 2 & 1 & 1 \\
\hline               
\end{tabular}
\caption{Total amount of UEEs within one second in which at least one UEE occurs. This listing is not accumulative.}
\label{tab:uee1s}
\end{table}
\newpage

\begin{figure}[!h]
  \begin{center}
    	\includegraphics[width=0.7\textwidth]{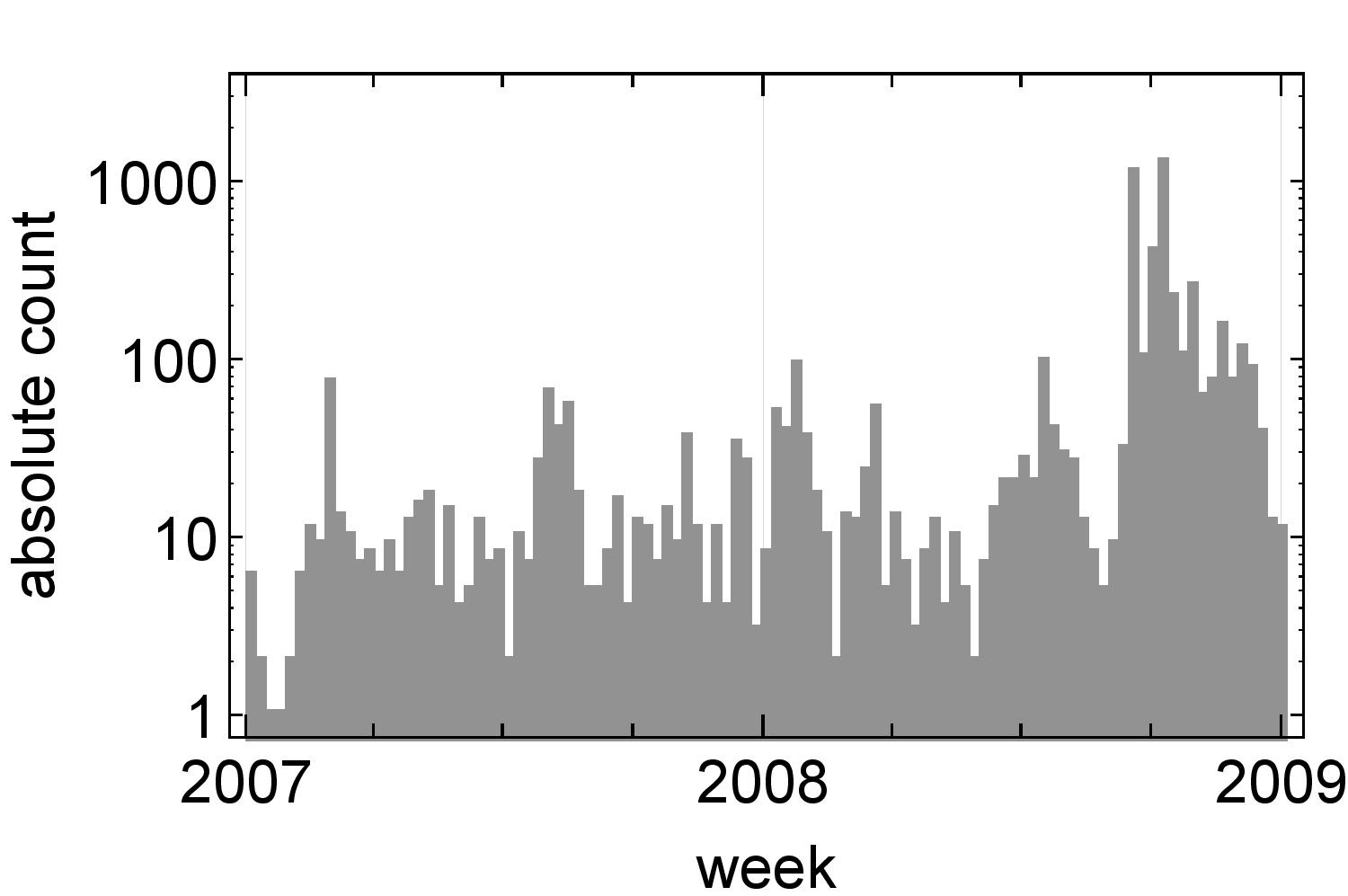}
  \end{center}
 \caption{Absolute count on a logarithmic scale of all UEE occurrences versus time with a bin size of one week. At the end of the third and at the beginning of the last quarter in 2008, the number of UEEs virtually explodes due to the financial crisis.}
 \label{fig:timeoccuee}
\end{figure}

The absolute count of all UEE occurrences is shown in Fig.~\ref{fig:timeoccuee}. We see an enormous increase of UEEs at the beginning of the financial crisis in September and October 2008, $46.5\,\%$ of the observed UEEs are flash crashes, hence $53.5\,\%$ of them are flash spikes. These observations corroborate the results of Golub \textit{et al.}~\cite{golub2011mini, golub2012high}, who used a similar data set. To determine the UEE size, \textit{i.e.} the price deviation between the beginning and the end of an UEE, we have to define what the end of an UEE is.

\begin{figure}[!h]
  \begin{center}
    	\includegraphics[width=0.7\textwidth]{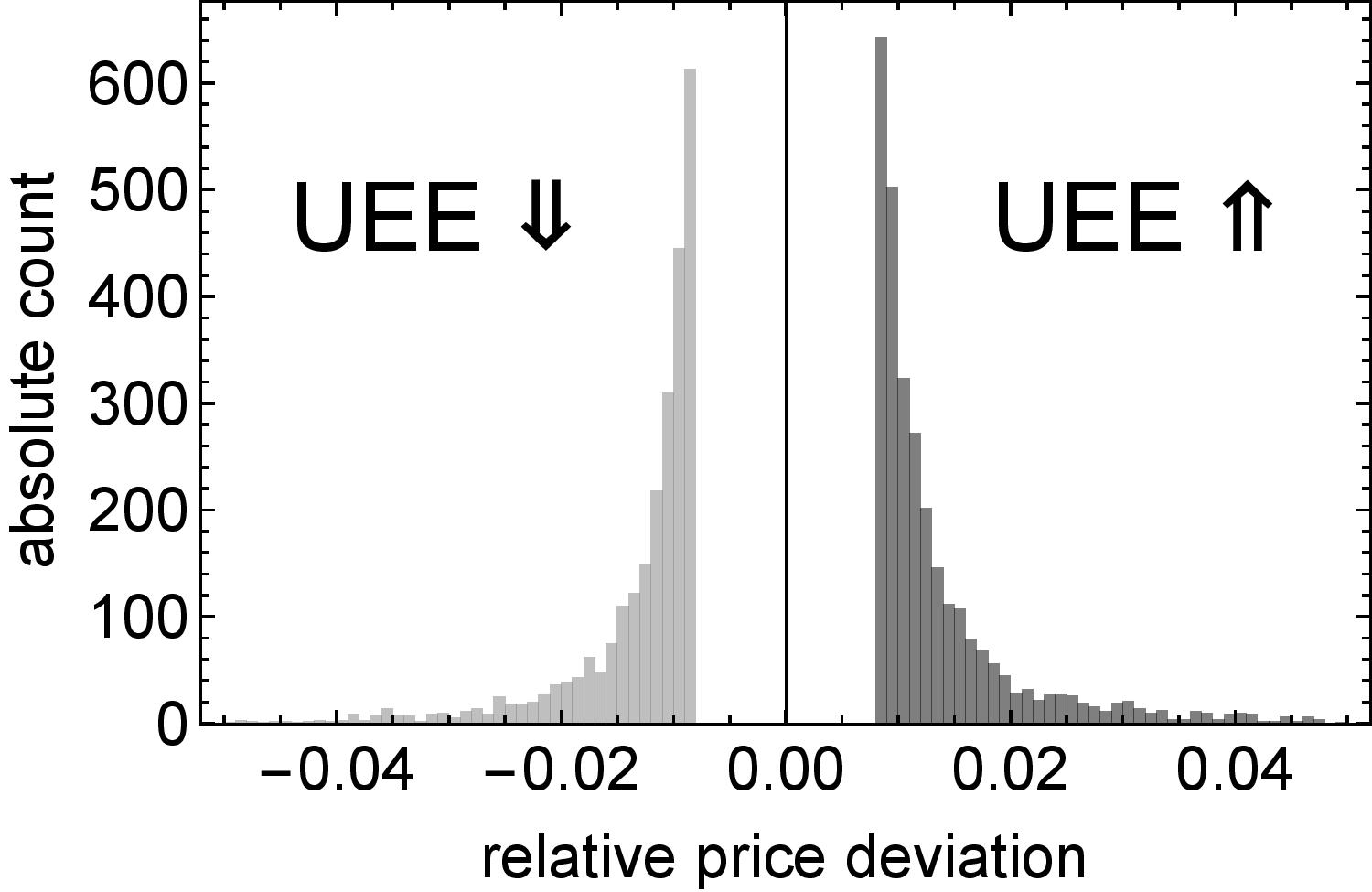}
  \end{center}
 \caption{Absolute count of all UEE occurrences versus the relative price deviations on a linear scale with a bin size of $0.001$. The gap around zero is due to the UEE criterion. The double arrows indicate the direction of the price change.}
 \label{fig:ueesize}
\end{figure}

In our study the first trade after an UEE either reverses the price trend or occurs after an at least one second lasting trading pause, whichever happens earlier. The former trigger applies in $78.3\,\%$ of all cases, whereas the latter ratio is $21.7\,\%$. The UEE size histogram in Fig.~\ref{fig:ueesize} clearly reflects the $8\,\permil$ UEE criterion. As an average relative price jump we calculate $-13.9\,\permil$ and $15.3\,\permil$, respectively, and the probability for a larger price jump than $5\,\%$ is $1.1\,\%$ for flash crashes and $1.68\,\%$ for flash spikes.

\newpage
\subsection{Mechanism for UEEs}
\label{sec4}

We consider two opposing scenarios that generate UEEs. First, if --- in case of a flash crash (spike) --- the best bid (ask) during an UEE decreases (increases) abruptly from one time step to the next, a large sell (buy) market order has to be the reason for this extreme event.
Second and in contrast, if the corresponding best value changes in minor price jumps, small market orders have created this UEE. Since these events occur within a few milliseconds, only the first scenario can be caused by a human trader, whereas many minor market orders within a very short time interval can only be placed by high frequent trading algorithms.
In this context, we exclude a random ``coherent'' synchronization by human traders induced by an external information that could theoretically also lead to the same scenario.

\begin{figure}[!h]
\begin{minipage}{.49\textwidth}
\includegraphics[width=1\textwidth]{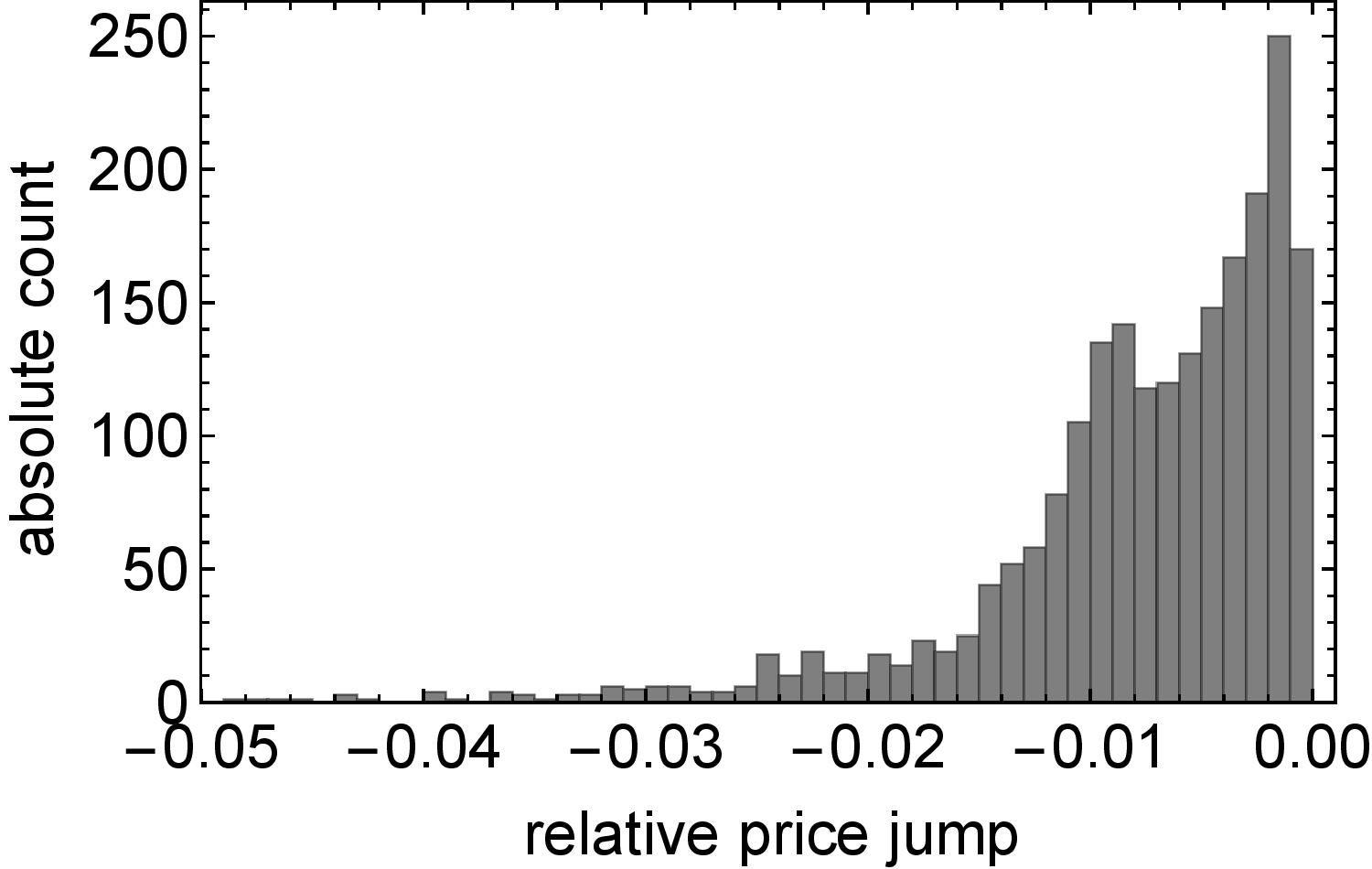}
\end{minipage}
\hfill
\begin{minipage}{.49\textwidth}
\includegraphics[width=1\textwidth]{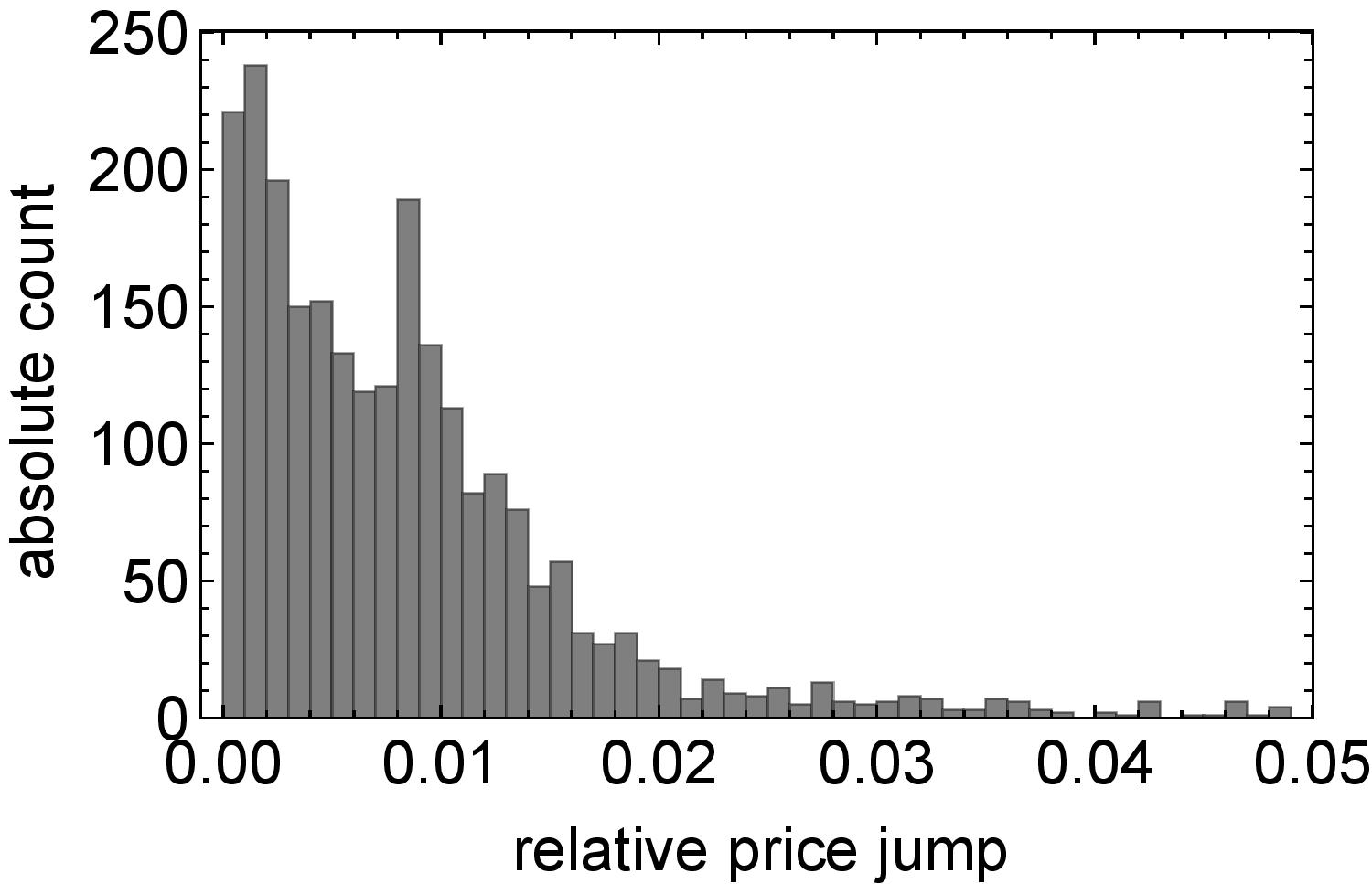}
\end{minipage}
\caption{Absolute count of all flash crash (spike) occurrences on a linear scale each represented by its largest best bid (ask) price jump.}
 \label{fig:bb_crash}
\end{figure}

To determine which effect dominates, we calculate for each UEE the largest price jump of the best bid or the best ask, respectively. The results are shown in Fig.~\ref{fig:bb_crash}. In $57\,\%$ ($60\,\%$) of all flash crash (spike) events there is one market order that causes a return of at least $0.5\,\%$. Moreover, in $40\,\%$ ($45\,\%$) of the cases, a single market order leads to a price jump that is big enough to fulfill the $8\,\permil$ UEE criterion. This shows that huge market orders play a major role in the emergence of UEEs. Hence all traders, not only high frequent traders, are able to induce an UEE. Furthermore these findings can be interpreted in context of the results of Golub \textit{et al.}\cite{golub2012high}: with a similar dataset these authors find $67.85\%$ of all UEEs to be initiated by so called Intermarket Sweep Orders (ISOs). Although we are not able to identify ISOs this observation supports our conclusion that large market orders are the main trigger for UEEs since ISOs typically result in one single huge trade event.

Which stocks are vulnerable to UEEs? --- We compare (not shown) typical order book snapshots of GS (Goldman Sachs Group Inc.) and AAPL (Apple Inc.). GS has 124 UEEs in 2008 and AAPL only has nine. We see that the order book of GS has many empty price levels whereas the order book of AAPL is densely populated. A large market order would therefore lead to a larger price jump for GS than the same market order would do for AAPL. We conclude that UEEs are rare for AAPL because the stock is very liquid and market orders would have to be very big and costly to generate UEEs.

\subsection{Impact and recovery of UEEs}
\label{sec5}

\begin{figure}[!h]
  \begin{center}
    	\includegraphics[width=.52\textwidth]{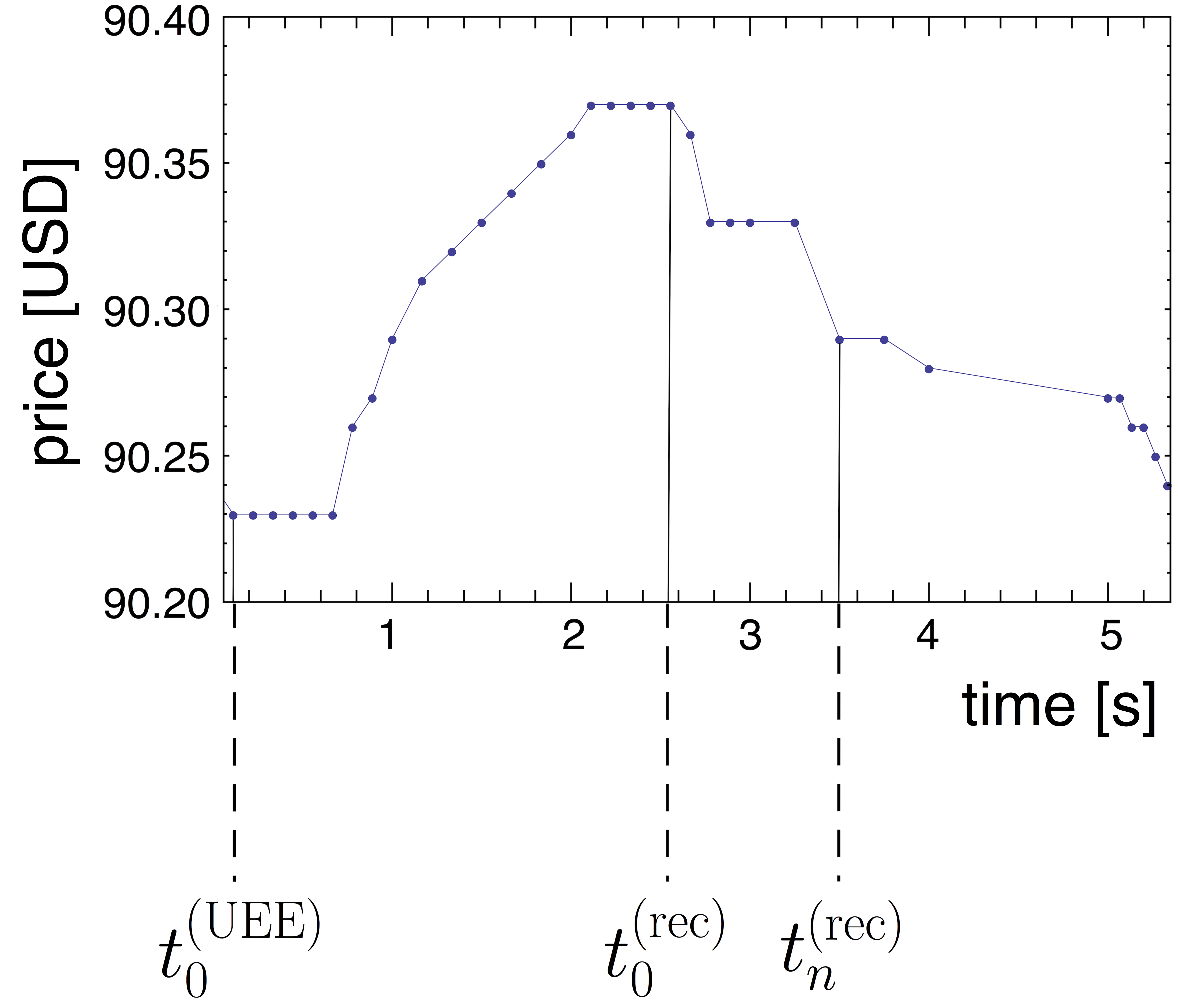}
  \end{center}
 \caption{Price time series of a typical UEE (AAPL, NASDAQ, 01.12.2008) to illustrate the definition of the recovery rate $\eta_n$. The vertical lines mark the specific instances $t_{\mathsmaller{0}}^{\mathsmaller{\mathrm{(UEE)}}}$, $t_{\mathsmaller{0}}^{\mathsmaller{\mathrm{(rec)}}}$ and $t_{\mathsmaller{n}}^{\mathsmaller{\mathrm{(rec)}}}$ in time.}
\label{fig:definitions_pdf}
\end{figure}

\begin{figure}[!h]
\begin{minipage}{.45\textwidth}
   \includegraphics[width=1\textwidth]{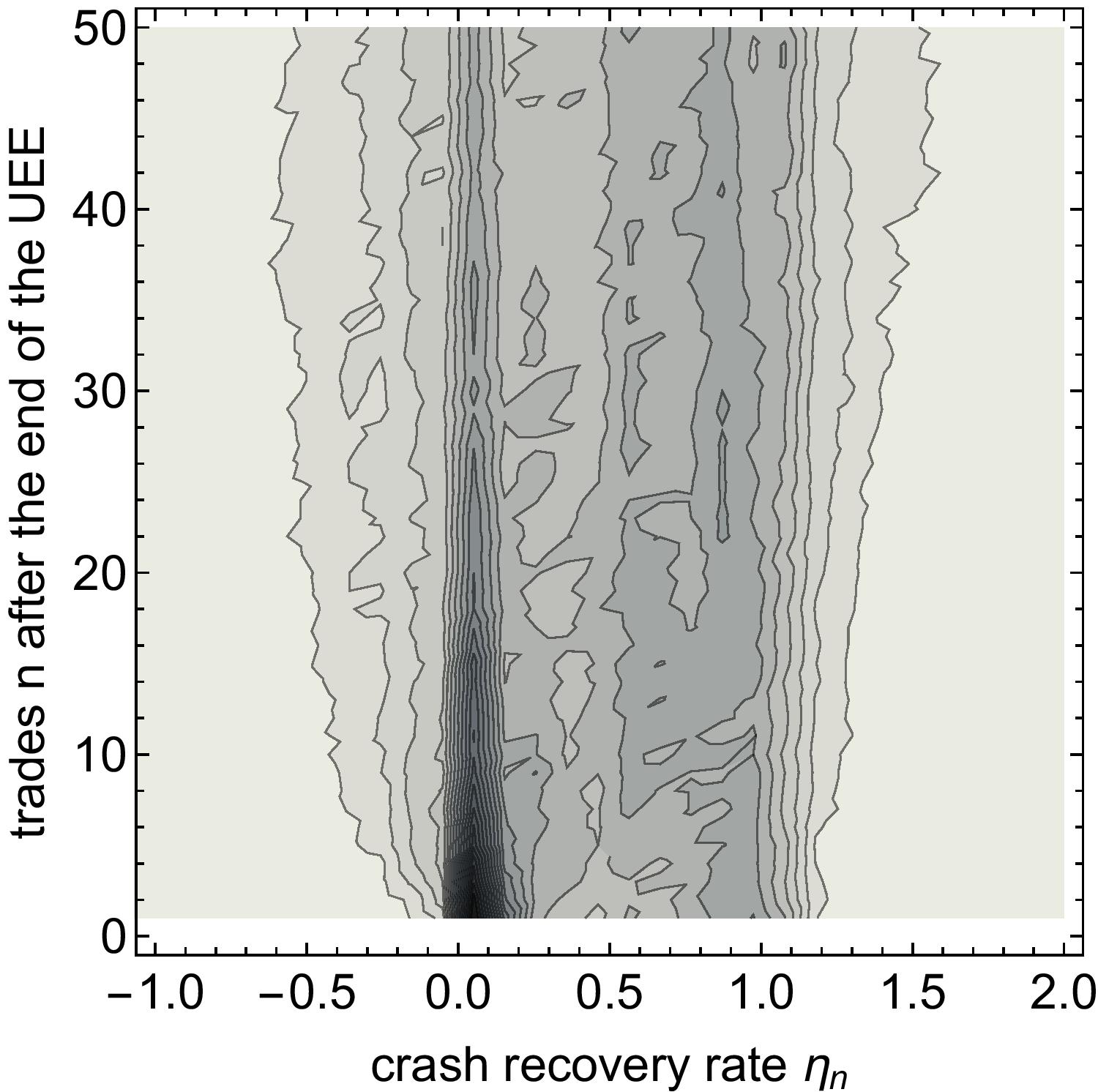}
\end{minipage}
\hfill
\begin{minipage}{.55\textwidth}
    \includegraphics[width=1\textwidth]{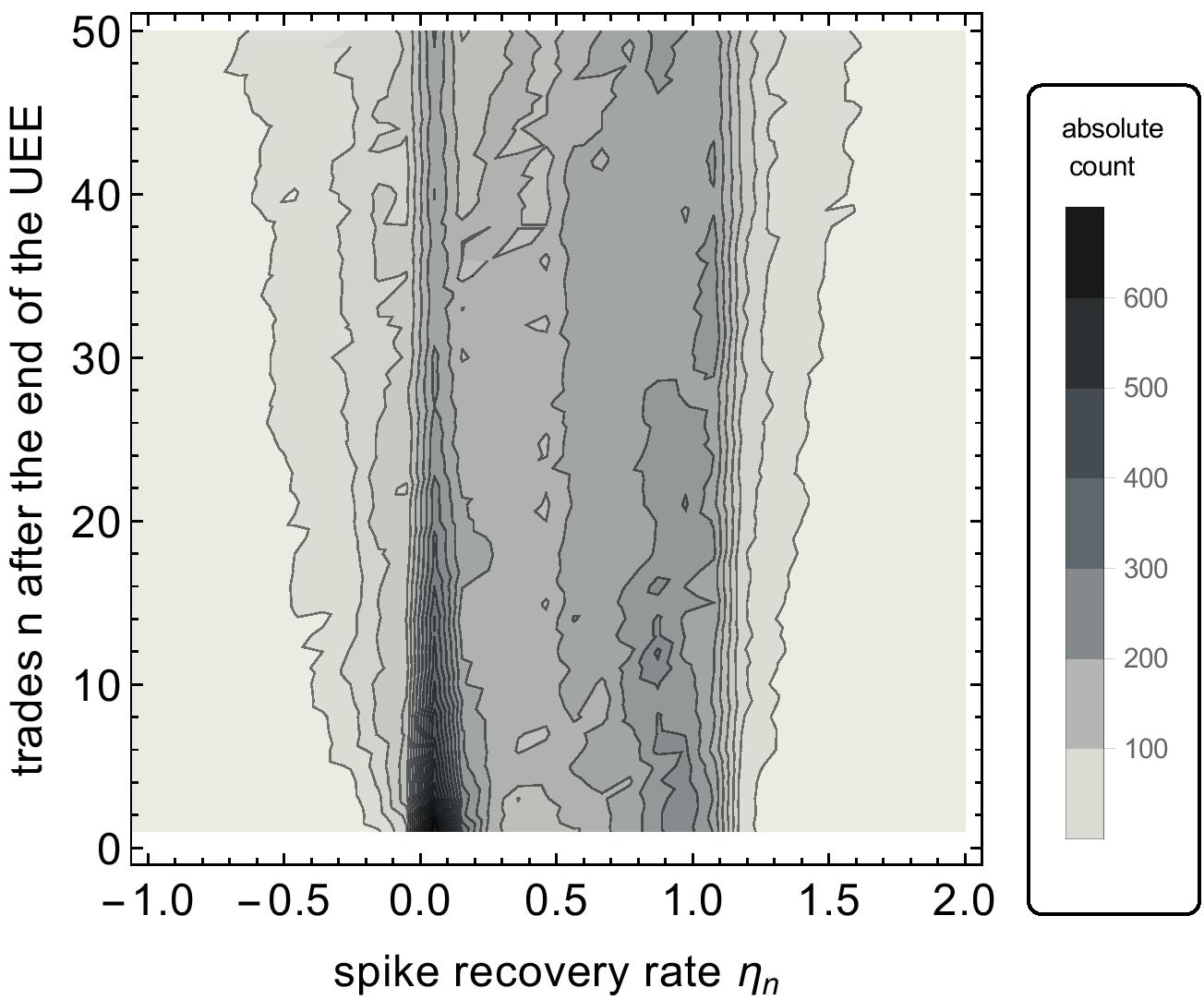}
\end{minipage}
 \caption{Absolute count  of all flash crashes (spikes) as level curves over the plane of recovery rate $\eta_n$ and number of trades $n$ (with $1 \le n \le 50$) after the end of the UEE. (Due to better visualization an empty line is added for $n=0$.)}
 \label{fig:rec_crash_pdf}
\end{figure}

A further step to a better understanding of UEEs is the observation of the price behavior after the UEEs' have reached their extreme values. For every UEE we calculate the crash/spike recovery rate $\eta_n$. As illustrated in Fig. \ref{fig:definitions_pdf}, we denote by $t_{\mathsmaller{0}}^{\mathsmaller{\mathrm{(UEE)}}}$ the time at which the UEE sets in, by $t_{\mathsmaller{0}}^{\mathsmaller{\mathrm{(rec)}}}$ the time at which the recovery sets in and by $t_{\mathsmaller{n}}^{\mathsmaller{\mathrm{(rec)}}}$ the time at which the $n$-th trade after beginning of the recovery occurs. The corresponding stock prices are $S(t_{\mathsmaller{0}}^{\mathsmaller{\mathrm{(UEE)}}})$, $S(t_{\mathsmaller{0}}^{\mathsmaller{\mathrm{(rec)}}})$ and $S(t_{\mathsmaller{n}}^{\mathsmaller{\mathrm{(rec)}}})$. We introduce the crash/spike recovery rate by the definition
\begin{align}
 \eta_n \ = \ \frac{S(t_{\mathsmaller{0}}^{\mathsmaller{\mathrm{(rec)}}}) - S(t_{\mathsmaller{n}}^{\mathsmaller{\mathrm{(rec)}}})}{S(t_{\mathsmaller{0}}^{\mathsmaller{\mathrm{(rec)}}})-S(t_{\mathsmaller{0}}^{\mathsmaller{\mathrm{(UEE)}}})}\: .
\end{align}
Thus, $\eta_n$ describes how much the UEE has recovered up to trade $n$, $\eta_n=0$ means the stock price has not recovered and is the same as $S(t_{\mathsmaller{0}}^{\mathsmaller{\mathrm{(rec)}}})$ whereas $\eta_n=1$ indicates that there was a full recovery and the stock price is equal to $S(t_{\mathsmaller{0}}^{\mathsmaller{\mathrm{(UEE)}}})$. Of course, $\eta_n$ can be greater than one or less than zero as well, because $S(t_{\mathsmaller{0}}^{\mathsmaller{\mathrm{(rec)}}})$ and $S(t_{\mathsmaller{0}}^{\mathsmaller{\mathrm{(UEE)}}})$ are not subject to boundaries. 
\newpage
Figure~\ref{fig:rec_crash_pdf} shows the histograms for all crashes or spikes, respectively, and $1 \le n \le 50$. While in both cases it is most likely that the stock price stays at the UEE extremum directly after the event, there are also cases in which it recovers immediately. Over the next trades, the histogram blurs considerably which is plausible as trading continues. Large price recovery even after a few trades is in sharp contrast with long lasting price impacts of market orders, see Bouchaud \textit{et al.}\cite{bouchaud2004fluctuations}. Even the impact of more complex market influences, as for example effective market orders consisting of many smaller market orders, are known to have long lasting market impacts \cite{toth2011anomalous}.

\begin{figure}[!h]
\begin{minipage}{.49\textwidth}
    	\includegraphics[width=1\textwidth]{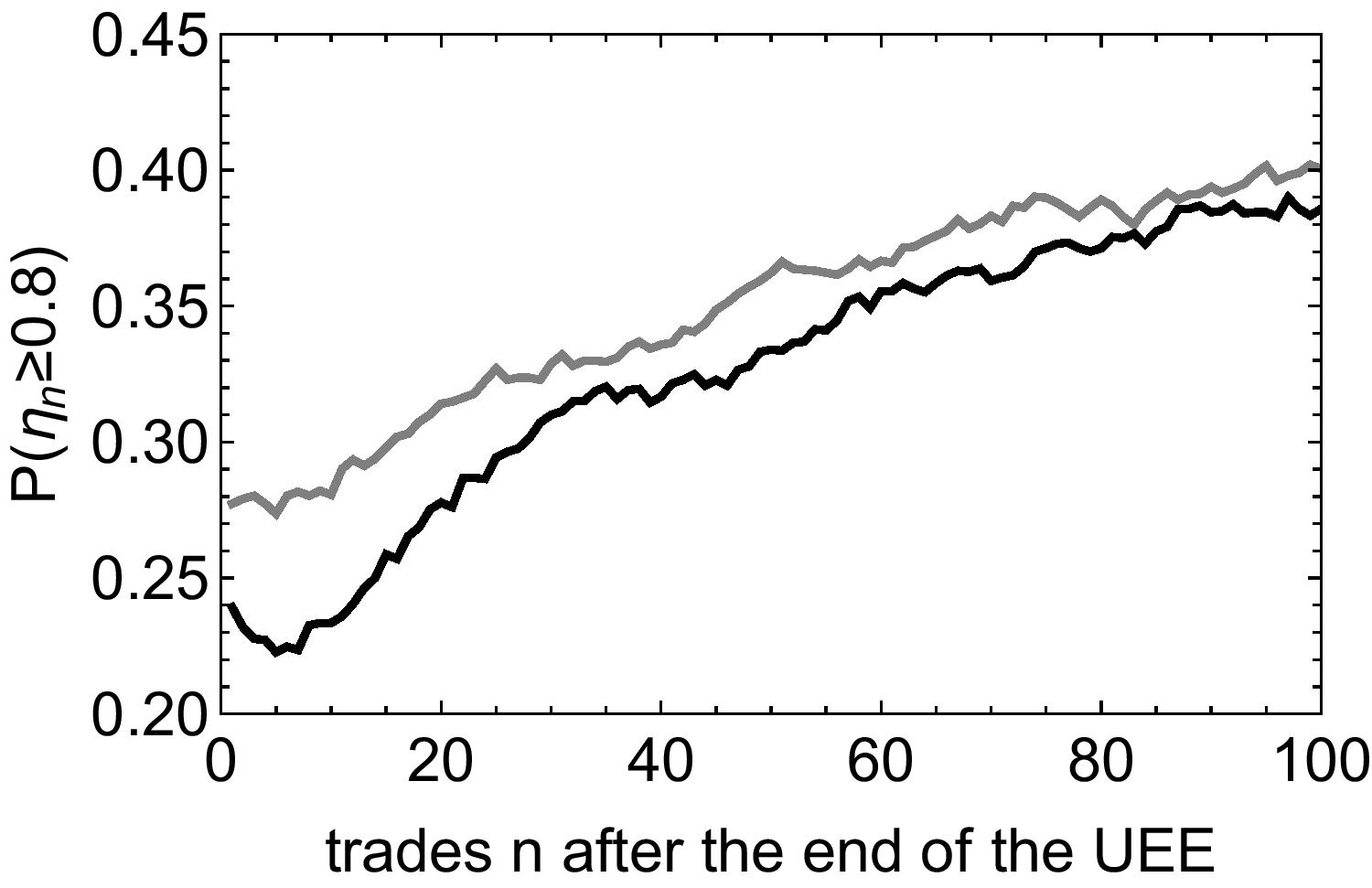}
\end{minipage}
\hfill
\begin{minipage}{.49\textwidth}
    	\includegraphics[width=1\textwidth]{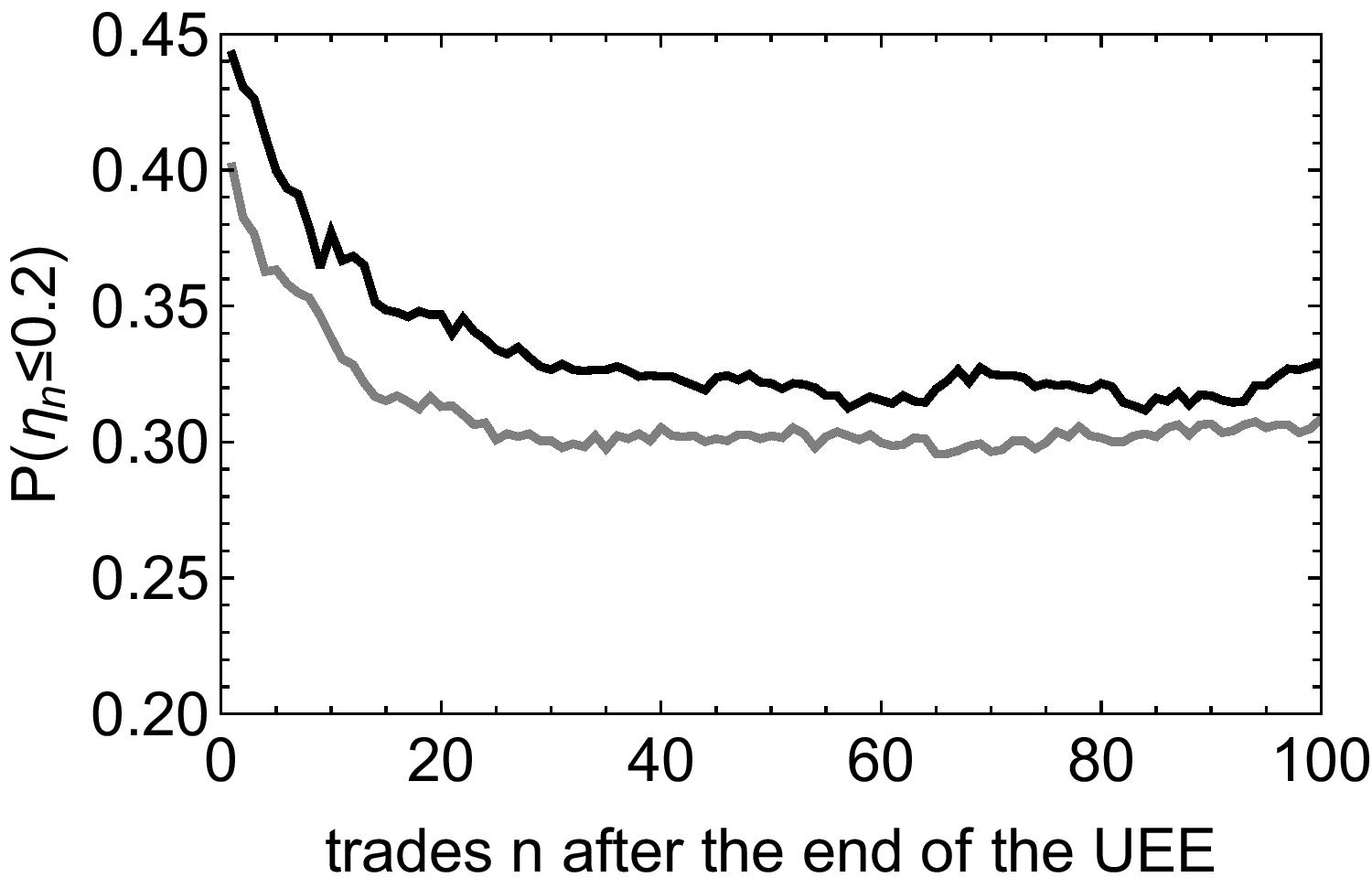}
\end{minipage}
 \caption{Probabilities $P(\eta_n \ge 0.8)$ and $P(\eta_n \le 0.2)$ for the event ``UEE is recovered by at least $80\,\%$ or $20\,\%$, respectively'' versus the number $n$ of trades after the end of the UEE for flash crashes (black) and flash spikes (gray).}
 \label{fig:rec}
\end{figure}

Furthermore we also calculate the probability for $\eta_n \ge 0.8$ and $\eta_n \le 0.2$ crash/spike recovery rates as function of $n$ to answer the question: How likely is an UEE to almost recover or tendentially remain at their extremum? The results are shown in Fig.~\ref{fig:rec}. We see that as more trades go by, it becomes more likely that an UEE recovers by at least $80\,\%$. In contrast the probability of not recovering decreases until $n=30$, but then constantly about $30\,\%$ of these UEEs recover by less than $20\,\%$.
Unexpectedly, the amount of recovered flash crashes decreases for about five trades before it starts to increase monotonously. This indicates that in about at least $2\%$ of the UEEs a flash crash is closely spaced by downward price trends that might be considered as ``aftershocks'' of the respective extreme event. This does not seem to be the case for the flash spikes.

\section{Discussion}
\label{sec6}
We found a total number of 5529 UEEs in 2007 and 2008. The financial sector clearly dominated with an average of 33.35 UEEs per company. Our analysis supports the observation that stocks with lower liquidity are more likely to generate large price changes which can lead to UEEs.
Differentiating between flash crashes and flash spikes, the probability to observe the latter is higher by $7\%$. Concerning the frequency of UEEs it is  not uncommon to find more than one UEE per second. In fact the probability for this to occur is approximately $9\%$. 
Especially in times of financial crisis the total frequency of UEEs surges as also found by Johnson \textit{et al.}\cite{Johnson2013}.

To analyse the mechanism that leads to the occurence of UEEs we distinguished between two microscopic interpretations. On the one hand the scenario put forward by Johnson \textit{et al.}\cite{Johnson2013} includes a new \textit{all-machine phase} in which many small market orders occur that --- considering the time scale --- can only be linked to HFT. On the other hand we assumed that already large market orders would be able to cause the observed price changes within a typical UEE, in agreement with Golub \textit{et al.}\cite{golub2012high}. This would conversely indicate that not only algorithmic but also human traders could trigger UEEs. In our analysis this interpretation is substantiated by the fact that about $60\%$ of the UEEs contain one market order that already generates a return of $0.5\%$. In contrast to the first scenario that considers UEEs as completely driven by HFT, the observation that large market orders play a major role shows that this has not necessarily to be the case. Thus, our analysis does not corroborate the conclusions of Johnson \textit{et al.}\cite{Johnson2013}. Nevertheless, it is worthwhile mentioning that the time resolution in our analysis was limited to one second. It is thus desirable to carry out such an analysis with millisecond accuracy.

Having investigated possible causes of UEEs we studied their impact on the subsequent trading. As one would anticipate the amount of recovering UEEs rises with the total number of trades ahead. Nevertheless a fraction of $30\%$ of all UEEs recover less than $20\%$ with respect to the price level before the UEE occurs. 

Regarding the fact that about $25\%$ of the UEEs are almost recovered after one trade, there must be processes in the order book dynamics that lead to this observation. At this point one could argue that this should be linked to different types of UEEs with respect to their impact on the following trades which one could try to classify, as an extension to already existing allocations as suggested by Nokerman\cite{nokermantaxonomy}. The fast price recovery indicates that flash crashes do not reflect the perceptions of investors, but are rather short lasting accidents which are rapidly repaired, see also  Christensen \textit{et al.}\cite{christensen2014fact}. Interesting further questions in this direction are the long term price impact, the impact of a flash crash in one stock on the price of other stocks \cite{wang2016cross}, and aftershocks \cite{petersen2010market}. 

\section*{Acknowledgements}

We thank Rudi Sch\"afer and Thilo Schmitt for fruitful discussions and helpful advice at the initial stages of the project.


\end{document}